\documentclass[a4paper, pra, twocolumn,superscriptaddress,showpacs,english]{revtex4-1}
\usepackage[T1]{fontenc}
\usepackage[latin9]{inputenc}
\usepackage{float}
\usepackage{amssymb}
\usepackage{amsmath}
\usepackage{graphicx}
\usepackage{babel}
\usepackage{color}
\def\cp#1{\mathbf{#1}}

\begin{document}
	
	\title{Non-universal Fermi polaron in quasi two-dimensional quantum gases}
	
	\author{Yue-Ran Shi}
	\affiliation{Department of Physics, Renmin University of China, Beijing 100872, China}
	\author{Jin-Ge Chen}
	\affiliation{Department of Physics, Renmin University of China, Beijing 100872, China}
	\author{Kuiyi Gao}
	\email{kgao@ruc.edu.cn}
	\affiliation{Department of Physics, Renmin University of China, Beijing 100872, China}
	\author{Wei Zhang}
	\email{wzhangl@ruc.edu.cn}
	\affiliation{Department of Physics, Renmin University of China, Beijing 100872, China}
	\affiliation{Beijing Academy of Quantum Information Sciences, Beijing 100193, China}
	\affiliation{Beijing Key Laboratory of Opto-electronic Functional Materials and Micro-nano Devices, Renmin University of China, Beijing 100872,China}

	\begin{abstract}
		We consider an impurity problem in a quasi-two-dimensional Fermi gas,  where a spin-down impurity is immersed in a Fermi sea of $N$ spin-up atoms. Using a variational approach and an effective two-channel model, we obtain the energies of the system for a wide range of interaction strength and for various different mass ratios between the impurity and the background fermion in the context of heteronuclear mixture. It is demonstrated that in a quasi-two-dimensional Fermi gas there exists a transition of the ground state from polaron in the weakly interacting region to molecule in the strongly interacting region. The critical interaction strength of the polaron--molecule transition is non-universal and depends on the particle density of the background Fermi sea. We also investigate the properties of the excited repulsive polaron state, and find similar non-universal behavior.
	\end{abstract}
	\pacs{}
	
	\maketitle
	
	\section{Introduction}
	\label{sec:intro}
	
	Ultracold atomic gases have been widely used as excellent platforms for the study of quantum many-body phenomena due to its high tunability in the past several decades~\cite{Dalfovo, Stringari, Bloch08}. In particular, it is of great interest to investigate low-dimensional systems of quantum gases and explore rich physics of many exotic phases of quantum matters, such as Tonks-Girardeau gas~\cite{Paredes, Kinoshita}, Fulde-Farrell-Larkin-Ovchinnikov (FFLO) state~\cite{Liao}, Berezinskii-Kosterlitz-Thouless (BKT) phase~\cite{Hadzibabic06, Murthy14}, pseudogap phase~\cite{Feld, Murthy18}, and so on. Low-dimensional systems of quantum gases can be realized by trapping the laser cooled atoms in various types of potentials, including atom chips~\cite{Fortagh}, light sheet~\cite{Vale}, TEM01 trap~\cite{Vale2}, surface trap~\cite{Greiner} and optical lattices~\cite{Paredes, Kinoshita, Liao, Hadzibabic06, Murthy14, Feld, Murthy18, Kohl0506, Jochim}. For fermions, Feshbach resonance~\cite{Inouye98, mfr, chin10, ofr} further provide exciting possibilities to create low-dimensional ultracold Fermi gases with a tunable inter-particle interaction, therefore allow for exploration of strongly correlated fermionic matters writh the interplay of interaction and quantum fluctuation~\cite{Levinsen14, Turlapov17}. It is natural to consider a Fermi gas trapped in a deep one-dimensional optical lattice, where most of the atoms are assumed to be in the lowest energy level of the trapping potential. As a result, we can achieve a quasi-two-dimensional (quasi-2D) Fermi system~\cite{quasi2D1, quasi2D2, quasi2D3, Feld, quasi2D5, quasi2D6}, in which the effect of the strong trapping potential can be modeled by a one-dimensional harmonic potential $V(z) = \frac{1}{2}m\omega_z^2z^2$. The temperature $T$ and the Fermi energy $E_F=\hbar^2k_F^2/2m$ also satisfy the conditions for quasi-two-dimensionality as $k_BT \ll E_F \ll \hbar \omega_z $~\cite{tightconfined}, where $k_F$ is the Fermi vector and $m$ the atomic mass. However. these are not always strictly satisfied in experiments, especially when the interaction between the Fermions increases~\cite{Dyke16}.
	
	The advent of two-dimensional Fermi gas has greatly renewed interest in the so-called polaron problem, which can be considered as the limiting case of a spin-imbalanced Fermi gas. It is of great importance to understand the nature of the fate of impurities as the interaction varies by tuning the Feshbach resonances. A natural expectation is the the system would undergo a transition and change its statistics as the impurity binds with background fermions~\cite{polaron1, polaron2}. Indeed, if the interaction is sufficiently weak, the impurity is dressed by the density fluctuation of the Fermi sea and hence forms a polaron state~\cite{polaronstate1, polaronstate2}. With increasing attractive interaction, the impurity binds a background particle and forms a molecule~\cite{moleculestate}. 
	Some previous studies show that a transition between polaron and molecule states exists both in theoretical calculations~\cite{2Dpolarontheory1, 2Dpolarontheory2, 2Dpolarontheory3} and experiments~\cite{2DpolaronMichael}. However, the significant population of the exited transverse mode of the harmonic potential in the strongly interacting regime~\cite{Kestner06, effctiveH, Zhang08} leads to non-universal situation of this phase transition, which might provide the key to understand the discrepancy of the transition points between theoretical prediction and experimental outcome in this quasi-2D systems. 
	
	In this work, we investigate the quasi-2D polaron problem using a variational approach and an effective two-channel model, in which the highly excited transversal levels and the resulting non-universal properties are taken into account by introducing a phenomenological degree of freedom of dressed molecules. We specifically consider two different kinds of background fermions, $^6$Li and $^{40}$K, as have been well investigated in experiments\cite{2DpolaronMichael, 2DpolaronZwierlein, 2DpolaronThomas, 2DpolaronGrimm}. Therefore three different mass ratios $\eta = m_\downarrow/m_\uparrow = 1, 6.64$ and $1/6.64$ are considered.  The main purpose of this work is to compare the difference from the single-channel model and the two-channel model, and also to prove that the value of $\hbar\omega_z/E_F$ strongly affects the position of the polaron-molecule transition. We numerically calculate the transition points for different mass ratios, and compare our results with previous experiments. We show the results of polaron energy, the wave function fraction for the polaron state, and the effective mass are sensitively dependent on particle density and mass ratio. Furthermore, we discuss the excited repulsive polaron state in this system and find similar non-universal behavior.
	
\section{Formalism}
\label{sec:form}
	\begin{figure}[tbp]
		\centering{}
		\includegraphics[width=0.98\columnwidth]{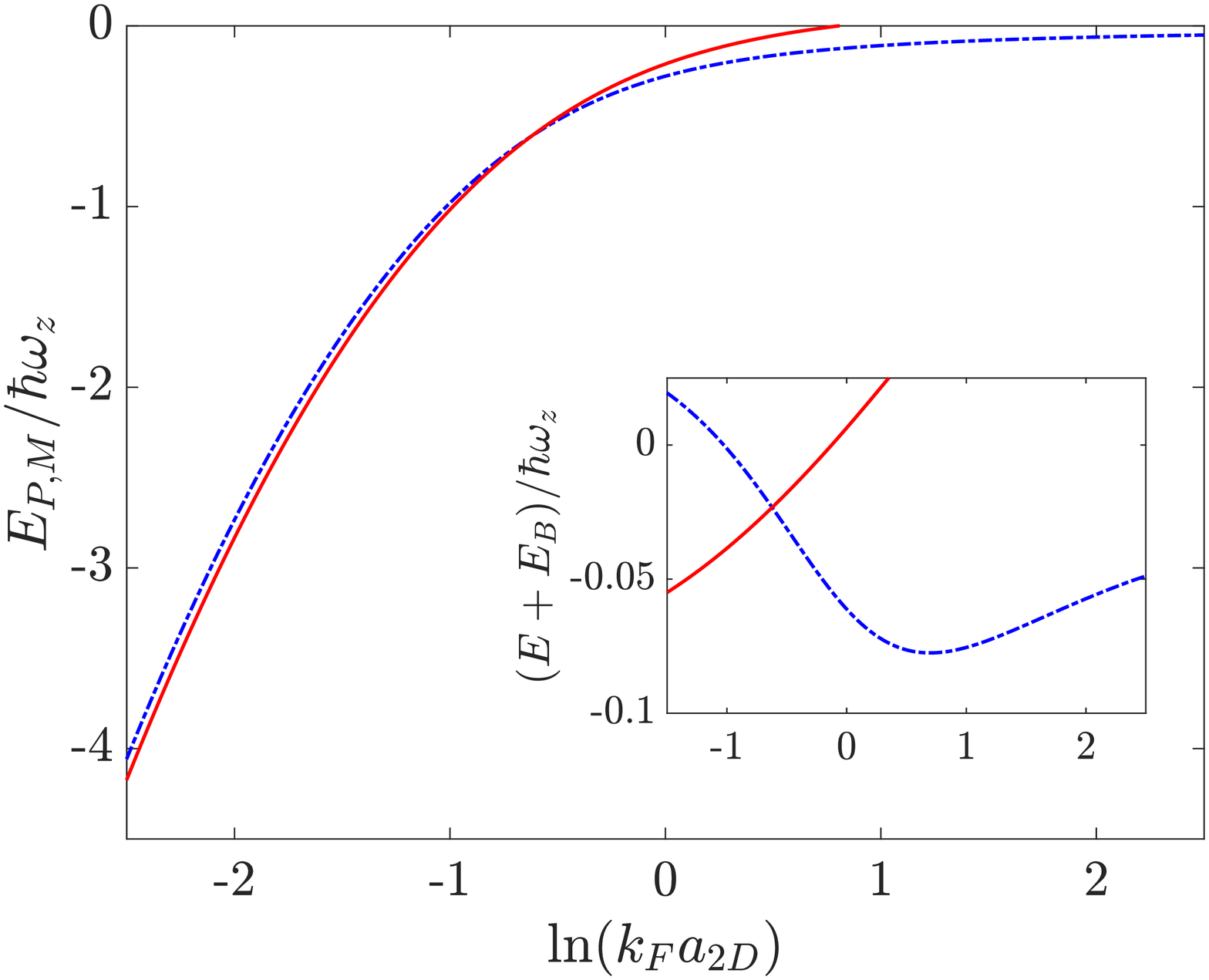}
		\caption{(Color online) Energies of  the molecule (red solid line) and the attractive polaron (blue dotdashed line) states with mass ratios $\eta =1$. Here we use the parameters for $^{40}K$ atoms and fix $\hbar \omega_z/E_F = 7.85$, same as~\cite{2DpolaronMichael}. The inset shows the result of $E+E_B$, and the transition point here is $\ln(k_Fa_{2D}) = -0.6224$.}
		\label{fig:energy}
	\end{figure}
	We consider an atomic Fermi gas trapped in a $(3-D)$-dimensional harmonic potential with an isotropic trapping frequency of $\omega_z$ in the transversal directions. Although in the following discussion we will focus on the quasi-2D case with $D=2$ only, in this section we present a general formalism which is valid for both cases of $D=1$ and $2$. Besides, while it is possible to use a single-channel approach with a pseudo-potential for a wide Feshbach resonance, for more general purposes we adopt the conventional two-channel field theory which is valid for arbitrary scenarios. The Hamiltonian thus can be written as~\cite{Bruun04,effctiveH}
	\begin{eqnarray}
		\label{exactH}
		H &=& \sum\limits_{\sigma = \uparrow, \downarrow} \int d^3{\bf r} \Psi_\sigma^\dagger \bigg(-\frac{1}{2}\nabla^2 + \frac{1}{2} \sum_{i = 1}^{3-D} x_i^2 \bigg)  \Psi_\sigma \nonumber\\
		&& +  \int d^3{\bf r} \Phi^\dagger \bigg(-\frac{1}{4}\nabla^2 + \sum_{i = 1}^{3-D} x_i^2  + \nu_b \bigg)  \Phi \nonumber\\
		&& + \ g_b \int  \int d^3{\bf r} (\psi_\uparrow^\dagger \psi_\downarrow^\dagger \Phi + H.c.)  \nonumber\\
		&& + \ U_b \int  \int d^3{\bf r}  \int d^3{\bf r} \psi_\uparrow^\dagger \psi_\downarrow^\dagger\psi_\downarrow\psi_\uparrow
	\end{eqnarray}
	with the atomic field operator $\Psi_\sigma({\bf r})$ ($\sigma = \uparrow, \downarrow$) and the molecular field operator $\Phi({\bf r})$. Here, the energy unit of $\hbar \omega_z$ is used, and the bare parameters ($\nu_b, g_b, U_b$) in the Hamiltonian are defined as follows: $\nu_b$ is the detuning, $g_b$ is the atom-molecule coupling constant, and $U_b$ is background atomic interaction constant. They are related to the corresponding physical parameters via the renormalization relations
	\begin{eqnarray}
		\label{renormalization}
		U_c^{-1}=\int \frac{d^3 {\bf k}}{(2\pi)^3} \frac{1}{2\epsilon_{\bf k}}, \ \Gamma^{-1}=1-U_pU_c^{-1}, \nonumber \\
		U_b=\Gamma U_p,  \ g_b=\Gamma g_p,  \ \nu_b=\nu_p + \Gamma g_p^2U_c^{-1},
	\end{eqnarray}
	with the subscript $p$ denoting the physical parameters and $\epsilon_{\bf k}=\hbar^2 {\bf k}^2/2m$. The integral in $U_c^{-1}$ is defined in three dimensions, with a cut off $k_c$. The physical parameters are defined from the parameters of the Feshbach resonance as $U_p=4\pi a_{bg}/a_t$, $g_p=\sqrt{4\pi \mu_{co}W|a_{bg}|/a_t \hbar \omega_z}$, and $\nu_p=\mu_{co}(B-B_0)/\hbar \omega_z$, where $\mu_{co}$ is the differential magnetic moment between the two channels and the $s$-wave scattering length near resonance is $a_s=a_{bg}[1-W/(B-B_0)]$ (with the background scattering length $a_{bg}$, the resonance width $W$ and center field of the  resonance $B_0$). 
	\begin{figure}[tbp]
		\centering{}
		\includegraphics[width=0.98\columnwidth]{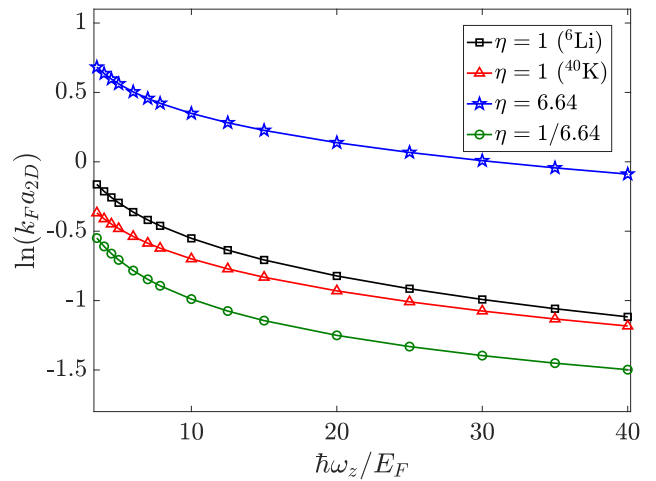}
		\caption{(Color online) The positions of trasition point vary with $\hbar\omega_z/E_F$. The black square and red triangle lines represent the equal mass case, where the atoms are $^6$Li and $^{40}$K, respectively. From bottom to top the mass ratio is increasing.} 
		\label{fig:transition}
	\end{figure}
	
	The Hamiltonian Eq.~(\ref{exactH}) captures the full ingredients of the trapping potential and the Feshbach resonance, however it can not be directly linked to the unique properties of low-dimensional physics, hence it is desirable to obtain an effective low-dimensional Hamiltonian if the many-body energy scales related to the density are much smaller than the trapping energy $\hbar \omega_z$. In this limit, one can obtain a Hamiltonian of the following form~\cite{Bruun04, effctiveH, moleculestate}
	\begin{eqnarray}
		\label{effH}
		H_{\text{eff}} &=& \sum\limits_{\bf k,\sigma} \epsilon_{\bf k,\sigma} c_{\bf k,\sigma}^\dagger c _ {\bf k,\sigma}
		+ \sum\limits_{\bf q} \bigg(\frac{1}{2} \epsilon_{\bf q} + \lambda_{b} \bigg) b_{\bf q}^\dagger b_{\bf q} \nonumber \\
		&&+ \frac{\alpha_b}{L^{D/2}} \sum\limits_{\bf k \bf q } \bigg(b_{\bf q}^\dagger c_{{\bf k}+ \frac{\bf q}{2}, \uparrow}  c_{-{\bf k}+ \frac{\bf q}{2}, \downarrow} + b_{\bf q} c_{-{\bf k}+ \frac{\bf q}{2}, \downarrow} ^\dagger  c_{{\bf k}+ \frac{\bf q}{2}, \uparrow} ^\dagger  \bigg) \nonumber \\
		&&+ \frac{V_b}{L^{D}} \sum\limits_{\bf k \bf k' \bf q}  c_{{\bf k}+ \frac{\bf q}{2}, \uparrow} ^\dagger c_{-{\bf k}+ \frac{\bf q}{2}, \downarrow} ^\dagger c_{{\bf k'}+\frac{\bf q}{2}, \downarrow} c_{-{\bf k'}+ \frac{\bf q}{2}, \uparrow},
	\end{eqnarray}
	where the relative detuning $\lambda_b$ (in units of $\hbar\omega_z$), the coupling constant $\alpha_b$ (in units of $\hbar \omega_z a^{D/2}_t$), and the background interaction in the open channel $V_b$ (in units of $\hbar \omega_z a_t^D$) are determined phenomenologically. Similarly, these bare parameters in $H_{\text{eff}}$ are related to the physical ones by 
	\begin{eqnarray}
		\label{renormalization2}
		V_c^{-1}=\int \frac{d^D {\bf k}}{(2\pi)^D} \frac{1}{2\epsilon_{\bf k} + 3 - D}, \ \Omega^{-1}=1-V_pV_c^{-1}, \nonumber \\
		V_p=\Omega^{-1} V_b,  \ \alpha_p=\Omega^{-1} \alpha_b,  \ \lambda_p=\lambda_b - \Omega \alpha_p^2V_c^{-1}.
	\end{eqnarray}
	%
	
	\begin{figure}[tbp]
		\centering{}
		\includegraphics[width=0.98\columnwidth]{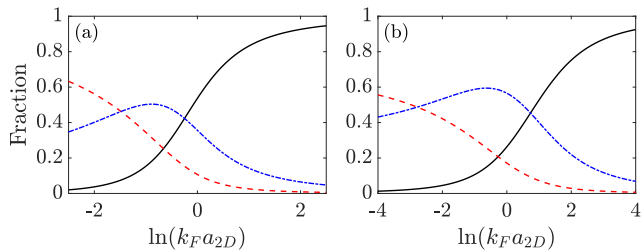}
		\caption{(Color online) The normalized contribution of polaron wave functions in the three different channels with (a) $\eta=1$ and (b) $\eta=6.64$. Here the three lines for $|\gamma|^2$ (solid black line), $\sum_{\bf q} |\alpha_{\bf q}|^2$ (dashed red line) and $\sum_{\bf k,q}|\beta_{\bf q}|^2$ (dashed-dotted blue line) represent the bare impurity, molecule and particle-hole excitation, respectively.}
		\label{fig:wavefunc}
	\end{figure}
	\begin{figure}[tbp]
		\centering{}
		\includegraphics[width=0.95\columnwidth]{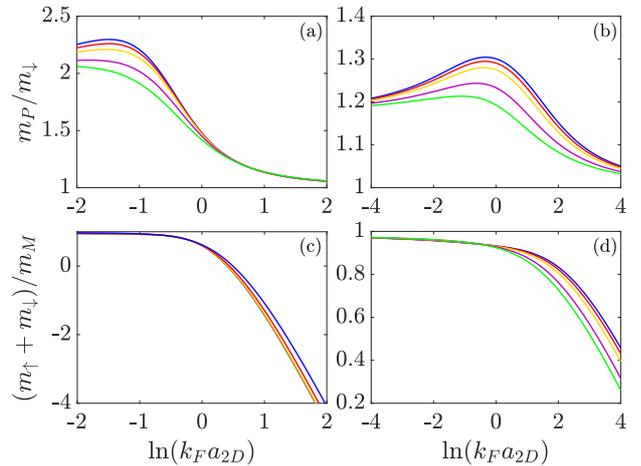}
		\caption{(Color online) Effective mass of polaron state with (a) $\eta = 1$, (b) $\eta = 6.64$ and inverse effective mass for molecule state with  (c) $\eta = 1$, (d) $\eta = 6.64$. The five lines from top to bottom represents $\hbar\omega_z/E_F = 40, 30, 20, 7.85$ and $3.5$, respectively.  }
		\label{fig:effmass}
	\end{figure}
	
	All these parameters in the effective two-channel model can be determined phenomenologically by matching the two-body bound state solved exactly from the original Hamiltonian Eq.~(\ref{exactH}).\cite{effctiveH}
	To compare with experiments, we focus on the two component Fermi gases of $^6$Li and $^{40}$K atoms, as well as their heteronuclear mixtures. The scattering parameters for $^6$Li ($^{40}$K) is $W \simeq 300$G (8G) , $a_{bg} \simeq -1405a_B$  $(174a_B)$, $\mu_{co}\simeq 2\mu_B$ $(1.68\mu_B)$~\cite{mfr, Bartenstein, chin10}. With a strong trap frequency of $\omega_z\simeq2\pi \times 40$ kHz, the physical parameters for the interaction are given by $g_p=310.2141$ $(26.1660)$ and $U_p=-4.5531$ $(1.4459)$. In the calculation below, one can use a common form of the renormalization relation for a quasi-2D model
	\begin{eqnarray}
		\label{renormalization3}
		\frac{1}{V_b-\frac{\alpha_b^2}{\lambda_b-x}} = \frac{1}{V_p-\frac{\alpha_p^2}{\lambda_p-x}} - \sum_{\cp k} \frac{1}{2\varepsilon_{\cp k} + (3-D)\hbar \omega_z},
	\end{eqnarray}
	where $\varepsilon_{\cp k}$ represents the two-dimensional dispersion.
	
	Usually this system has two possible ground states. For weak interaction, the ground state is expected to be a polaron state with an approximate wave function~\cite{polaronstate1}
	\begin{eqnarray}
		\label{eqn:polwf}
		|P({\bf Q})\rangle &=& \gamma
		c_{{\bf Q}\downarrow}^\dagger | {N} \rangle_\uparrow +\sum_{{\bf q}} \alpha_{\bf q} b^\dagger_{{\bf Q}+{\bf q}}c_{{\bf q}\uparrow}| {N} \rangle_\uparrow
		\nonumber \\
		&& +
		\sum_{\bf k,q} \beta_{{\bf k q}} c^\dagger_{{\bf Q}+{\bf q}-{\bf k}\downarrow} c^\dagger_{{\bf k}\uparrow} c_{{\bf q}\uparrow} | {N} \rangle_\uparrow,
	\end{eqnarray}
	where the first term is a bare impurity on top of the non-interacting majority Fermi sea $| {N} \rangle_\uparrow$, the second term represents a dressed molecule couples to the background Fermi sea to create a hole therein. The last term describes excitations containing one pair of particle-hole. Here, ${\bf Q}$ represents the center-of-mass momentum, and the summation of ${\bf k}$ is always over ${|{\bf k}|>k_F}$, and the summation of ${\bf q}$ is restricted below the Fermi level.
	

	The eigenenergy $E$ of the polaron state can be determined by minimizing $\langle P({\bf Q})|H-E|P({\bf Q})\rangle$, leading to the following equation
	\begin{eqnarray}
		\label{eqn:polE}
		E_P-\epsilon_{\bf Q \downarrow}=
		\sum_{\bf q}&\bigg[&\frac{1}{V_p - \frac{\alpha_p^2}{(\frac{1}{2}\epsilon_{\bf Q+q} + \lambda_{p}) +\epsilon_{\bf q \uparrow}-E_P}} \nonumber\\
		&&+ \sum_{\bf k}\frac{1}{E_{\bf Q k q}}
		- \sum_{\bf k} \frac{1}{2\epsilon_{\bf k} + \hbar \omega_z} \bigg]^{-1},
	\end{eqnarray}
	where $E_{\bf Q k q}=-E_P+\epsilon_{{\bf Q+q-k}\downarrow}+\epsilon_{{\bf k}\uparrow}-\epsilon_{{\bf q}\uparrow}$ and $E_P=E-\sum_{|{\bf k_0}| < k_F} \epsilon_{{\bf k_0}\uparrow}$.
	
	If the interaction strength is sufficiently strong, the impurity atom and one background particle prefer to form a molecule. We can also write a molecule wave function with one pair of particle-hole excitation as
	\begin{eqnarray}
		\label{eqn:molwf}
		|M({\bf Q})\rangle &=& \gamma b_{\bf Q}^\dagger| {N-1} \rangle_\uparrow + \sum\limits_{{\bf k}} \alpha_{\bf k} c_{{\bf Q - \bf k}\downarrow}^\dagger c_{{\bf k}\uparrow}^\dagger | {N-1} \rangle_\uparrow \nonumber \\
		&&+\sum_{\bf k, q} \beta^1_{\bf k q} b_{\bf Q + \bf q - \bf k}^\dagger c_{{\bf k}\uparrow}^\dagger c_{{\bf q}\uparrow}   | {N-1} \rangle_\uparrow \nonumber\\
		&&+
		\sum_{\bf k, k', q} \beta_{{\bf k k' q}}^2
		c^\dagger_{{\bf Q} + {\bf q} - {\bf k} - {\bf k'}\downarrow} c^\dagger_{{\bf k}\uparrow} c^\dagger_{{\bf k'}\uparrow}
		c_{{\bf q}\uparrow} | {N-1} \rangle_\uparrow. \nonumber
	\end{eqnarray}
	
	Note that the Fermi sea here only contains $N$-1 spin-up fermions, thus the threshold energy would be different from the polaron state. By minimizing $\langle M({\bf Q})|H-E|M({\bf Q})\rangle$, we obtain the molecule energy equations
	\begin{widetext}
		\begin{align}
			\label{eqn:molE}
			\bigg(  \frac{1}{V_p-\frac{\alpha_p^2}{\lambda_p-(E_M-\frac{1}{2} \epsilon_{\bf Q})}} &- \sum_{\bf k_1} \frac{1}{2\epsilon_{\bf k_1} + \hbar \omega_z} 
			+ \sum_{{\bf k'}} \frac{1}{E_{\bf Q k k' q}} \bigg) G({\bf k},{\bf q}) - \frac{1}{E_{\bf Q k}}   \sum_{\bf q_1} G({\bf k},{\bf q_1}) \\ \nonumber
			&-  \sum_{{\bf k'}}  \frac{G({\bf k'},{\bf q})}{E_{\bf Q k k' q}} + \frac{1}{E_{\bf k}} \cdot 
			\frac{\sum_{\bf k', q_1}\frac{G({\bf k'},{\bf q_1})}{E_{\bf Q k'}}}{\sum_{\bf k'} \frac{1}{E_{\bf Q k'}} - ( \frac{1}{V_p-\frac{\alpha_p^2}{\lambda_p-(E_M-\frac{1}{2} \epsilon_{\bf Q})}}  - \sum_{\bf k_1} \frac{1}{2\epsilon_{\bf k_1} + \hbar \omega_z})} =0,
		\end{align}
	\end{widetext}
	where $E_{\bf Q k} = E_M-\epsilon_{\bf Q-k\downarrow} - \epsilon_{{\bf k}\uparrow}-\sum_{|\bf q|<k_F}V_b$, $E_{\bf Q k k' q} = -E_M + \epsilon_{{\bf Q}+{\bf q}-{\bf k}-{\bf k'}\downarrow} + \epsilon_{{\bf k}\uparrow} + \epsilon_{{\bf k'}\uparrow} - \epsilon_{{\bf q}\uparrow}$ and $E_M$ is defined as $E-\sum_{|{\bf k_0}| < k_F} \epsilon_{{\bf k_0}\uparrow}$.
	
	
	\section{Results}
	\label{sec:res}
	Equations (\ref{eqn:polE}) and (\ref{eqn:molE}) can be solved numerically by discretizing the integrals to solvable matrix equations. The results for a zero center-of-mass momentum impurity are plotted in Fig.~\ref{fig:energy}. We first consider the case of mass ratio $\eta=m_\downarrow/m_\uparrow=1$ and $\hbar\omega_z /E_F = 7.85$, which is relevant to an experimentally feasible configuration where the impurity and background fermions are both $^{40}$K atoms. As we can see in Fig.~\ref{fig:energy}, there exists a polaron--molecule transition under these conditions. The transition takes place at $\ln(k_Fa_{\rm 2D}) = -0.6224$. Compared with the experimental measurement of $\ln(k_Fa_{\rm 2D}) = -0.4$ reported in Ref.~\cite{2DpolaronMichael}, our result is comparable with the pervious theoretical estimation of $\ln(k_Fa_{\rm 2D}) = -0.6$ using another version of quasi-2D model~\cite{2Dpolarontheory2}, and better than the outcome of $-0.8$ via a strictly two-dimensional theory~\cite{2Dpolarontheory1}.
	
	The most important finding of our treatment is the non-universality of the system. The polaron--molecule transition point $\ln(k_Fa_{\rm 2D})$ is not a universal constant, but depends on the density of background fermions and the trapping geometry. In Fig.~\ref{fig:transition} we show the dependence of transition point on $\hbar\omega_z/E_F$ for various mass ratio $\eta$, where the region below each transition line represents the molecular state regime for the corresponding case. For all mass ratios, the polaron--molecule transition is always present, and the transition point moves monotonically towards the strongly interacting limit by increasing $\hbar\omega_z/E_F$, leading to a narrower regime of molecular state. This trend can be understood by noticing that with tighter transversal confinement, the ground level population of the transversal trapping potential is enhanced, inducing a stronger fluctuation effect which tends to break the molecular state. We also notice that the transition line with heavier impurity, e.g., mass ratio $\eta = 6.64$, lies above the ones with lighter impurity, i.e., $\eta = 1$ and $1/6.64$. This is because the Fermi energy of lighter atoms is higher than that of heavier atoms for a given number density. And the polaron state is more favorable as the kinetic energy of background fermions surpasses the interaction energy binding the molecule state. 
	
	The non-universal behavior is of potential importance when comparing theoretical results and experimental measurements. In many cold atom experiments such as Ref.~\cite{2DpolaronMichael}, the quantum gases are trapped in a harmonic potential and the particle density varies from high to low by moving from the trap center to the edge. The response of the trapped gas under a global measurement is thus an average of all particles. Theoretical treatment can be performed within the local density approximation (LDA), under which the non-uniform system is considered as composed by segments with a local density fixed by the chemical potential with an on-site potential offset. The results shown in Fig.~\ref{fig:transition} suggest that the density dependence has to be taken into account when quantitative analysis is sought.
	
	\begin{figure}[tbp]
		\centering{}
		\includegraphics[width=0.98\columnwidth]{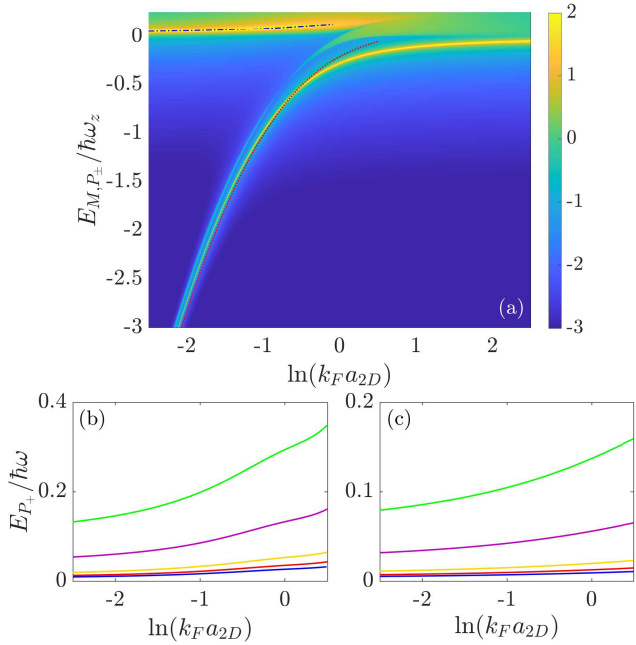}
		\caption{(a) Contour plot of the ($Log_{10}$ of the) spectral function  $A({\bf Q}=0, E_P)$ of the polaron state as a function of  $\ln(k_Fa_{2D})$ for a $^{40}$K system with $\hbar\omega_z /E_F = 7.85$. The red dotted line is the molecule energy obtained from Eq.(\ref{eqn:molE}).  On the BCS side, the repulsive polaron branch merges into the molecule-hole continuum which is denoted by the broad light area. (b), (c) Repulsive polaron energy $E_{P_+}$ with $\eta = 1$ (b) and $\eta = 6.64$ (c). The five lines from bottom to top represent $\hbar\omega_z/E_F = 40, 30, 20, 7.85$ and $3.5$, respectively. }
		\label{fig:spectralfunc}
	\end{figure}
	
	Another non-trivial finding of Fig.~\ref{fig:transition} is that the curves do not seem to saturate and remains non-universal for $\hbar \omega_z/E_F = 40$. This ratio corresponds to a very strong confinement and the quasi-2D condition should be satisfied quite well at the first glance. However, one must notice that in this system there is another energy scale characterizing the interaction energy, which can be chosen as the binding energy of the two-body bound state formed by the impurity and one background fermion. If the binding energy is comparable or even exceeds $\hbar \omega_z$, the excited levels of the transversal trapping potential would be significantly populated and the system remains non-universal. To better reveal this property, we further show the fraction of the bare impurity, the dressed molecule, and the two-dimensional fermions in the polaron wave function ansatz in Fig.~\ref{fig:wavefunc}. The three components are defined as
	\begin{eqnarray}
		\label{eqn:polfraction}
		&& |\gamma|^2=1- \sum_{\bf q} |\alpha_{\bf q}|^2-\sum_{\bf k, q} |\beta_{\bf k, q}|^2,\nonumber\\ 
		\sum_{\bf q} &&|\alpha_{\bf q}|^2 = \sum_{\bf q}  |\gamma|^2 \bigg[ \frac{\frac{\alpha_b}{E_{\bf q}}\cdot (V_b - \frac{\alpha_b^2}{E_{\bf q}})^{-1}}{ (V_b - \frac{\alpha_b^2}{E_{\bf q}})^{-1} + \sum_{\bf k} E_{\bf k, q}^{-1}}\bigg]^2, \nonumber\\
		\sum_{\bf k, q} &&|\beta_{\bf k, q}|^2 = \sum_{\bf k, q} |\gamma|^2 \bigg[ \frac{E_{\bf k,q}^{-1}}{ (V_b - \frac{\alpha_b^2}{E_{\bf q}})^{-1} + \sum_{\bf k_1} E_{\bf k_1, q}^{-1}} \bigg]^{2}.
	\end{eqnarray}
	In the weakling interacting limit, the dominant part of polaron wave function is the bare impurity part $|\gamma|^2$. In the intermediate and strong interaction regime, the dressed molecule fraction $\sum_{|q|}|\alpha_q|^2$ and fermionic contribution $\sum_{|k|,|q|}|\beta_q|^2$ are both significant. Reminding that the dressed molecule is a phenomenological description of the excited levels of the transversal trap, this observation suggests that the system is not strictly 2D yet for the range of interaction considered. 
	
	Next, we extend the discussion the finite center-of-mass momentum. The solution of Eqs.~(\ref{eqn:polE}) and (\ref{eqn:molE}) at small but finite momentum $|{\bf Q}| \ll k_{\uparrow F}$ can be expended like
	\begin{eqnarray}
		\label{effmass}
		E({\bf Q})=E({\bf 0})+ \frac{Q^2}{2m^*}
	\end{eqnarray}
	in both polaron and molecule state with $m^*$ the effective mass. In Fig.~\ref{fig:effmass} we show the effective mass for polaron and the inverse effective mass for molecule state. In the weakly interacting limit, the effective mass of the polaron state is the bare impurity mass $m^* \approx m_\downarrow$. With increasing interaction, the polaron state with zero center-of-mass momentum is always a metastable state with a positive effective mass. On the contrary, the molecule effective mass turns to be negative for weak enough interaction, and the molecule solution with $Q = 0$ becomes unstable. In the strongly interacting regime, the effective mass for both the polaron and molecule states approach the same limiting value $m^* \approx m_\uparrow+m_\downarrow$.
	
	Finally, we investigate the properties of the repulsive polaron branch with a positive energy, which was recently realized in experiments in both three dimensions~\cite{repulsivepol} and two dimensions~\cite{2Drepulsivepol1,2Drepulsivepol2}. We first write down the self-energy of a polaron state
	\begin{eqnarray}
		\label{selfenergy}
		\Sigma({\bf Q}, E_P) &=&
		\sum_{\bf q} \bigg[ \frac{1}{U_p - \frac{g_p^2}{(\frac{1}{2}\epsilon_{\bf Q+q} + \nu_{b}) +\epsilon_{\bf q \downarrow}-E_P}} \nonumber\\
		&+& \sum_{\bf k}\frac{1}{E_{\bf Q k q}-i0^+}
		- \sum_{\bf k} \frac{1}{2\epsilon_{\bf k} + \hbar \omega_z} \bigg]^{-1},
	\end{eqnarray}
	where the function $E_{\bf Q k q}$ is defined in Eq.~(\ref{eqn:polE}). The spectral function thus takes the form of
	\begin{eqnarray}
		\label{spectralfunc}
		A({\bf Q}, E_P) =  -2 \text{Im} \frac{1}{E_P+ i0^+ -\Sigma({\bf Q}, E_P)}.
	\end{eqnarray}
	
	In Fig.~\ref{fig:spectralfunc}(a), we plot the spectral function as a function of energy and interaction strength $\ln(k_Fa_{2D})$ for ${\bf Q}=0$ and $\eta=1$. There exist two branches where the spectral function is strongly peaked and labeled by bright yellow color. The lower branch corresponds to the attractive polaron state, which intersects with the molecular state (red dotted line) at the polaron--molecule transition. The thick upper branch labels the repulsive polaron, which is a fairly well defined excited state in the strongly interacting regime with a positive energy. As moving towards the weakly interacting regime, the repulsive polaron branch merges into the molecule--hole continuum. At small momenta, the energy of the two branches of polaron can also be calculated from
	\begin{eqnarray}
		\label{reppolenergy}
		E_P =  \text{Re}[\Sigma({\bf Q}=0, E_P)],
	\end{eqnarray}
	which is also displayed as black dashed-dotted line in Fig.~\ref{fig:spectralfunc}(a). We also observe a non-universal density dependence of the repulsive polaron energy, as demonstrated in Figs.~\ref{fig:spectralfunc}(b) and \ref{fig:spectralfunc}(c) for mass ratio $\eta = 1$ and $6.64$, respectively. 

	\section{Conclusion}
	\label{sec:con}
	We study the impurity problem in a background of fermionic particles trapped in a quasi-two dimensional confinement. Using an effective two-channel model, which incorporates the effect of excited states along the strongly confined direction, we investigate the various properties of the polaron and molecule states, including the eigenenergy, wave function, effective mass, and the spectral function. Since the transversal trapping potential for a quasi-two-dimensional configuration is finite, the system acquires another energy scale of $\hbar \omega_z$, and hence breaks the universality which is present in a strictly two-dimensional Fermi gas with contact interaction. Specifically, we find that the polaron--molecule transition point shows a strong density dependence on $\hbar\omega_z / E_F$ for all mass ratios. This observation if of particular interest when analyzing experimental results of cold atomic gases trapped in harmonic potentials, where the density is non-uniform.
	
	\acknowledgments
	We thank support from the National Key R\&D Program of China (Grant No.~2018YFA0306501), the National Natural Science Foundation of China (Grants No.~11522436, No.~11774425, No.~12074428), and the Beijing Natural Science Foundation (Grant No.~Z180013).
	

\end{document}